\documentclass[aps,prl,twocolumn,groupedaddress,showpacs]{revtex4}
\usepackage{bm}
\usepackage{graphicx}

\begin{document}

\title{Fermiology of Strongly Spin-Orbit Coupled Superconductor Sn$_{1-x}$In$_x$Te and its Implication to Topological Superconductivity}
\author{T. Sato,$^1$ Y. Tanaka,$^1$ K. Nakayama,$^1$ S. Souma,$^2$ T. Takahashi,$^{1,2}$ S. Sasaki,$^3$ Z. Ren,$^3$ A. A. Taskin,$^3$ Kouji Segawa,$^3$ and Yoichi Ando$^3$}
\affiliation{$^1$Department of Physics, Tohoku University, Sendai 980-8578, Japan}
\affiliation{$^2$WPI Research Center, Advanced Institute for Materials Research,
Tohoku University, Sendai 980-8577, Japan
}
\affiliation{$^3$Institute of Scientific and Industrial Research, Osaka University, Ibaraki, Osaka 567-0047, Japan}

\date{\today}

\begin{abstract}
We have performed angle-resolved photoemission spectroscopy of the strongly spin-orbit coupled low-carrier density superconductor Sn$_{1-x}$In$_x$Te ($x$ = 0.045) to elucidate the electronic states relevant to the possible occurrence of topological superconductivity recently reported for this compound from point-contact spectroscopy. The obtained energy-band structure reveals a small holelike Fermi surface centered at the L point of the bulk Brillouin zone, together with a signature of a topological surface state which indicates that this superconductor is essentially a doped topological crystalline insulator characterized by band inversion and mirror symmetry. A comparison of the electronic states with a band-non-inverted superconductor possessing a similar Fermi surface structure, Pb$_{1-x}$Tl$_x$Te, suggests that the anomalous behavior in the superconducting state of Sn$_{1-x}$In$_x$Te is likely to be related to the peculiar orbital characteristics of the bulk valence band and/or the presence of a topological surface state.
\end{abstract}

\maketitle
  
     Topological insulators (TIs) are a novel quantum state of matter where the bulk is an insulator with an ``inverted'' energy gap induced by a strong spin-orbit coupling (SOC), which leads to the emergence of unusual gapless edge/surface states protected by time-reversal symmetry \cite{HasanReview, SCZhangReview}. The discovery of TIs triggered the search for their superconducting (SC) analogues, topological superconductors (TSCs) \cite{SCZhangReview}. TSCs are accompanied by gapless Andreev bound states (ABSs) at the edge or surface \cite{RyuPRB,Volovik,Read,Qi,Sato}, which characterize the nontrivial topology of the bulk state and are often comprised of Majorana fermions \cite{WilczekNP}. Majorana fermions have been discussed to have a potential for fault-tolerant topological quantum computing \cite{AliceaRPP, BeenakkerArXiv} owing to their peculiar characteristics that particle is its own antiparticle \cite{WilczekNP}. The first plausible example of a TSC preserving time-reversal symmetry was copper-doped bismuth selenide (Cu$_x$Bi$_2$Se$_3$) \cite{HorPRL, FuBErgPRL, KrienerPRL, SasakiCuPRL}, where helical Majorana fermions are predicted to emerge \cite{FuBErgPRL}. The electronic properties of Cu$_x$Bi$_2$Se$_3$ have been intensively investigated by transport and spectroscopic studies including angle-resolved photoemission spectroscopy (ARPES) \cite{Kanigel, Bay, KrienerPRB2, HasanCuNP, TanakaCuPRB}, but it has been difficult to elucidate the nature of its superconducting state, partly because of the intrinsic inhomogeneity of this system \cite{KrienerPRB}.
     
   Very recently, it was reported that indium-doped tin telluride (Sn$_{1-x}$In$_x$Te, called In-SnTe here), a low-carrier density superconductor based on a narrow-gap IV-VI semiconductor, is possibly a new type of TSC, through the observation of a pronounced zero-bias conductance peak (ZBCP) in point-contact spectra below the SC transition temperature ($T_{\rm c}$) indicative of an unconventional surface ABS \cite{SasakiInPRL}. In contrast, a similar low-carrier density superconductor, thallium-doped lead telluride (Pb$_{1-x}$Tl$_x$Te, called Tl-PbTe here), appears to have no surface ABS and is likely to be non-topological \cite{SasakiInPRL}. These results are particularly intriguing in view of the fact that SnTe was recently elucidated to be materializing a new type of topological state called topological crystalline insulator (TCI) \cite{FuTCIPRL} associated with topologically-protected surface states that emerge due to a combination of band inversion and mirror symmetry \cite{TanakaNP, FuTCINC}, while PbTe is not a TCI \cite{TanakaNP}. It is thus of great importance to experimentally establish the Fermi surface (FS) and the band dispersions of In-SnTe and Tl-PbTe for both the bulk and the surface to clarify the electronic states relevant to the possible occurrence of topological superconductivity.
   
\begin{figure*}
\includegraphics[width=6.8in]{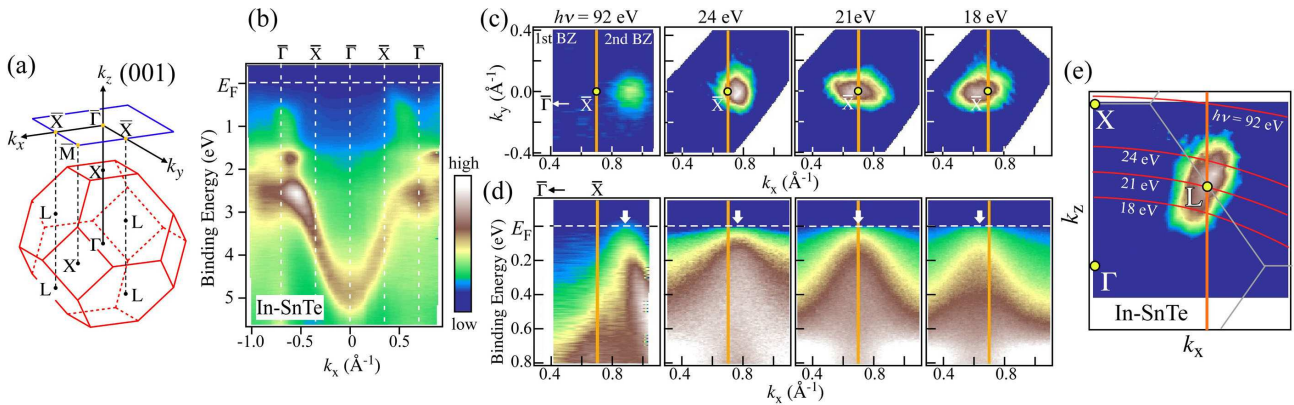}
\vspace{-0.4cm}
\caption{(Color online): (a) Bulk fcc BZ and corresponding tetragonal (001) surface BZ. (b) Valence-band ARPES intensity as a function of $k_x$ and binding energy ($E_{\rm B}$) for In-SnTe ($x$ = 0.045) on the (001) cleaved surface measured with $h\nu$ = 92 eV at $T$ = 30 K. (c) ARPES intensity mapping at $E_{\rm F}$ plotted as a function of in-plane wave vector ($k_x$ and $k_y$) measured with various photon energies; this intensity is obtained by integrating the spectra within $\pm$10 meV of $E_{\rm F}$. (d) Corresponding near-$E_{\rm F}$ band dispersion along the $\bar{\Gamma}$$\bar{X}$ cut. {\bf k} location of the local maxima of the valence band is indicated by a white arrow in (d). (e) ARPES intensity mapping at $E_{\rm F}$ plotted as a function of $k_x$ and $k_z$ obtained from the measurement by sweeping the photon energy.  Measured {\bf k} cuts in the $k_x$-$k_z$ plane for different photon energies are also indicated by red curves. Inner-potential value $V_0$ is 8.5 eV, which is the same as pristine SnTe \cite{TanakaNP}.
}
\end{figure*}

\begin{figure}
\includegraphics[width=3.0in]{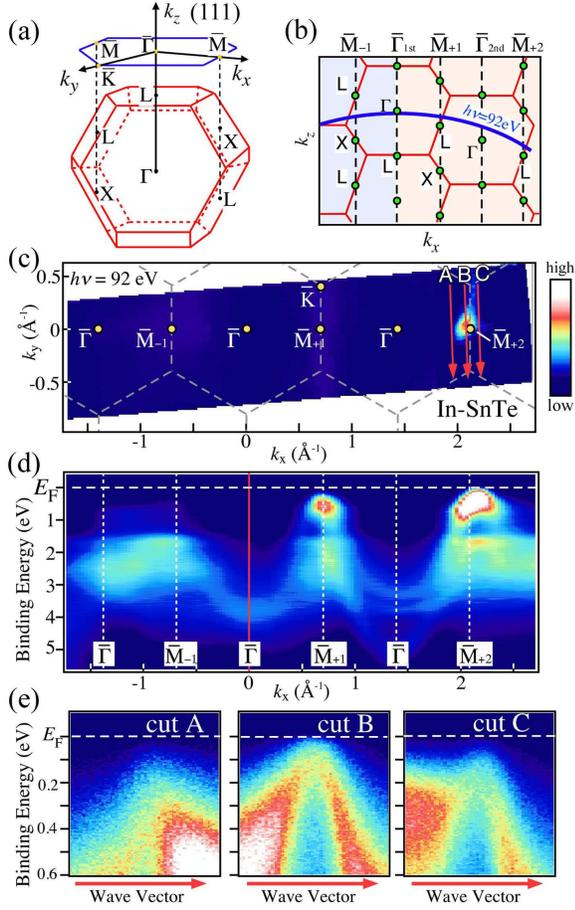}
\vspace{-0.4cm}
\caption{(Color online): (a) Bulk fcc BZ and corresponding hexagonal (111) surface BZ. (b) Bulk BZ in the $k_x$-$k_z$ plane together with a measured cut for $h\nu$ = 92 eV (blue curve). (c) ARPES intensity mapping at $E_{\rm F}$ covering a wide {\bf k} region for In-SnTe plotted as a function of in-plane wave vector measured at $h\nu$ = 92 eV on the (111)-cleaved surface. (d) Valence-band ARPES intensity as a function of $k_x$ and $E_{\rm B}$.  (e) Near-$E_{\rm F}$ ARPES intensity as a function of wave vector and $E_{\rm B}$ along representative {\bf k} cuts (A-C) in the surface BZ shown by red arrows in (c).}
\end{figure}

\begin{figure}
\includegraphics[width=3.4in]{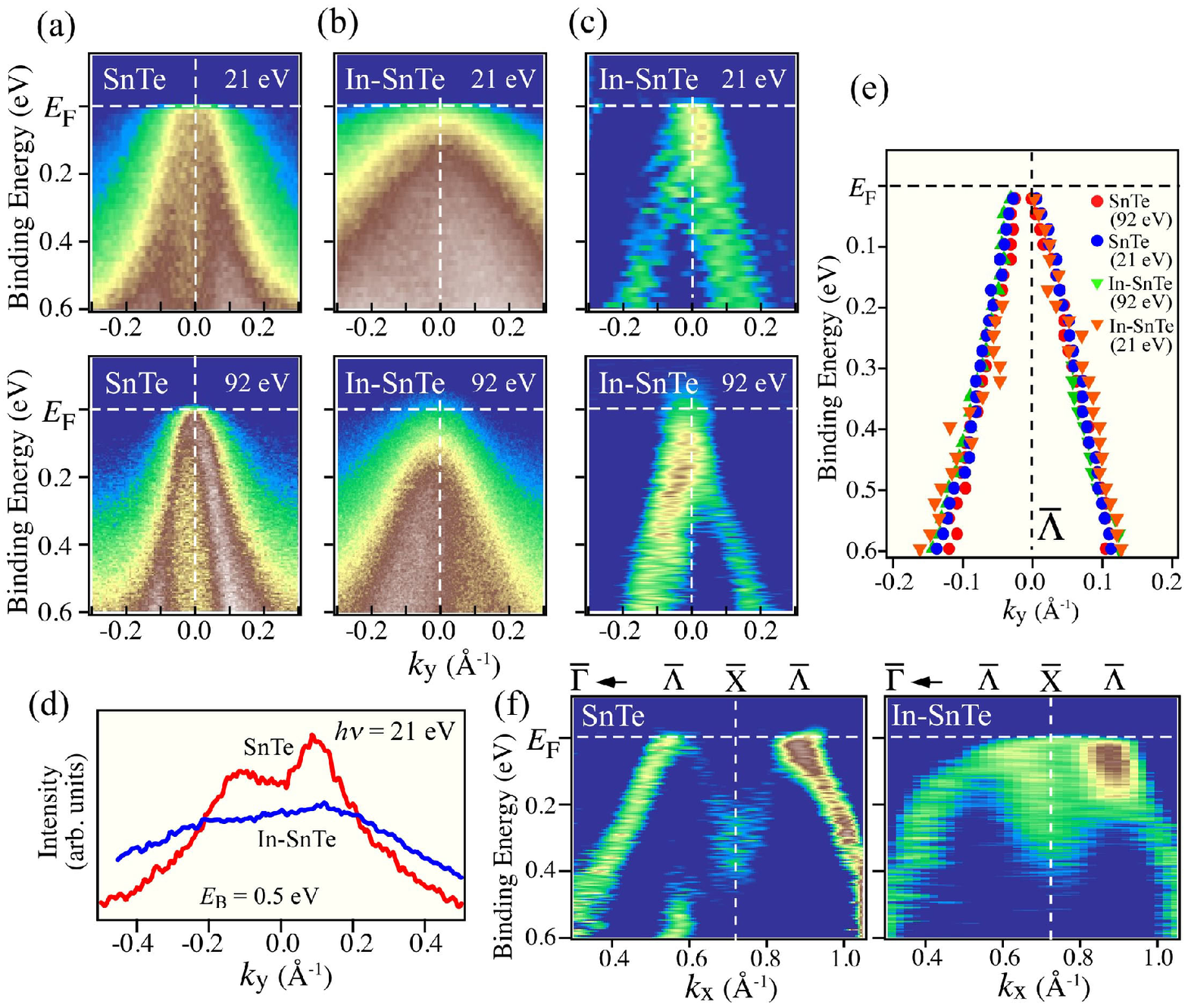}
\vspace{-0.4cm}
\caption{(Color online): (a) and (b), Comparison of near-$E_{\rm F}$ intensity as a function of $k_y$ and $E_{\rm B}$ across the $\bar{\Lambda}$ point between (a) pristine SnTe and (b) In-SnTe measured with $h\nu$ = 21 eV (top) and 92 eV (bottom) for the (001)-cleaved surface. (c) Second derivative of the data in (b), which was taken along the $k_y$ direction. (d) Comparison of the MDC at $E_{\rm B}$ = 0.5 eV between pristine SnTe and In-SnTe for $h\nu$ = 21 eV. (e) Band dispersions for $h\nu$ = 21 and 92 eV extracted by tracing the peak position of the second derivatives of the MDCs for SnTe and In-SnTe. (f) Comparison of the band dispersion for $h\nu$ = 21 eV derived from the second derivatives of the MDCs along the $\bar{\Gamma}$$\bar{X}$ cut between pristine SnTe (left) and In-SnTe (right).
}
\end{figure}

   In this Letter, we report high-resolution ARPES studies of In-SnTe and Tl-PbTe.  Most importantly, we found that the topological surface state exists in In-SnTe, although it is significantly weakened/broadened compared to pristine SnTe due to strong quasiparticle scattering caused by In doping.  On the other hand, no surface state was found in Tl-PbTe, which strongly suggests that the valence band of In-SnTe is inverted relative to Tl-PbTe, as is the case of their parent compounds.  In addition, we were able to elucidate the bulk FS topology of In-SnTe for two different cleavage planes (111) and (001), which gives conclusive evidence that the bulk FS is located at L points of the bulk Brillouin zone (BZ).
   
    High-quality single crystals of In-SnTe and pristine SnTe were grown by the vapor transport and the modified Bridgeman methods, respectively. Details of the sample preparations were described elsewhere \cite{SasakiInPRL, TanakaNP}. The indium concentration was determined to be $x$ = 0.045 with inductively coupled plasma atomic emission spectroscopy and is consistent with the observed $T_{\rm c}$ value of 1.2 K \cite{EricksonPRB}.  ARPES measurements were performed with the MBS-A1 and VG-Scienta SES2002 electron analyzers with a high-intensity helium discharge lamp at Tohoku University and also with tunable synchrotron lights at the beamline BL-7U at UVSOR as well as at the beamline BL28A at Photon Factory (KEK). To excite photoelectrons, we used the He I$\alpha$ resonance line ($h\nu$ = 21.218 eV), linearly polarized lights of 12-40 eV, and the circularly polarized lights of 50-100 eV, at Tohoku University, UVSOR, and Photon Factory, respectively. The energy and angular resolutions were set at 10-30 meV and 0.2$^{\circ}$, respectively. Samples were cleaved {\it in-situ} along the (001) or (111) crystal plane in an ultrahigh vacuum of 1$\times$10$^{-10}$ Torr. A shiny mirror-like surface was obtained after cleaving the samples, confirming its high quality. The Fermi level ($E_{\rm F}$) of the samples was referenced to that of a gold film evaporated onto the sample holder.
    
   Figure 1 displays the FS and band dispersions of the (001)-cleaved surface measured at $T$ = 30 K with various $h\nu$ for In-SnTe. In this sample orientation, the surface BZ projected onto the (001) plane is tetragonal, and the surface  $\bar{X}$ point corresponds to the projection of the L point in the bulk BZ [Fig. 1(a)]. As shown in Fig. 1(b), we clearly observe multiple dispersive bands in the valence-band region derived from Te 5$p$ orbitals \cite{LittlewoodPRL}, and the features are consistent with the previous ARPES data of pristine SnTe \cite{TanakaNP}. We also find a relatively weak dispersive feature approaching $E_{\rm F}$ near the $\bar{X}$ point, which is attributed to the top of the valence band. To elucidate the FS topology in the three-dimensional (3D) {\bf k} (wave vector) space, we have mapped out the intensity at $E_{\rm F}$ as a function of the in-plane wave vector ($k_x$ and $k_y$) for various photon energies, and the result is shown in Fig. 1(c). One can see in the plot for $h\nu$ = 92 eV that there is a single intensity spot centered at a point slightly away from $\bar{X}$ in the second BZ (right-hand side), which originates from the holelike dispersion as seen in the near-$E_{\rm F}$ band dispersion along the $\bar{\Gamma}$$\bar{X}$ cut [Fig. 1(d)].
   
Upon varying the photon energy, the bright intensity center in the FS mapping systematically changes its {\bf k} location in the BZ. For instance, the intensity is peaked exactly at the $\bar{X}$ point for $h\nu$ = 21 eV, whereas the peak for $h\nu$ = 18 eV is located a little closer to the $\bar{\Gamma}$ point in the first BZ (left-hand side). Such a variation in the intensity profile is also reflected in the band dispersion along the $\bar{\Gamma}$$\bar{X}$ cut shown in Fig. 1(d) where the {\bf k} position of the band-top changes upon $h\nu$ variation, as is traced by arrows. This demonstrates a considerable $k_z$ (wave vector perpendicular to the surface) dispersion of the band, suggesting that the observed spectral feature is dominated by the bulk contribution. 
To elucidate the 3D FS topology in more detail, we have mapped in Fig. 1(e) the ARPES intensity at $E_{\rm F}$ in the $k_x$-$k_z$ plane, which clarifies that the FS is centered at the L point of the bulk BZ and has an ellipsoidal shape elongated along the $\Gamma$L direction (note that the longer axis of the ellipsoid appears to be slightly tilted with respect to the $\Gamma$L direction, but this is probably an artifact resulting from a finite $k_z$ broadening).  Such an elongated shape of the bulk FS well explains the observed change in the in-plane FS mapping and the band dispersion in Figs. 1(c) and (d). We thus conclude that the bulk FS of In-SnTe is composed of a hole pocket at the L point. This is consistent with the band calculation \cite{LittlewoodPRL} and the hole-doping nature of In substitution for Sn.

   To further elucidate the bulk FS topology of In-SnTe, we have also investigated the electronic states for the (111)-cleaved surface. The (111)-oriented samples for the ARPES measurements were selected by carefully picking up those pieces that present six-fold-symmetric x-ray Laue diffraction spots, while (001)-oriented samples can be easily distinguished with their four-fold-symmetric Laue spots. In the (111) geometry, the surface BZ is hexagonal, and the $\bar{\Gamma}$ ($\bar{M}$) point corresponds to the projection of the L and $\Gamma$ (X and L) points of the bulk BZ [Fig. 2(a)]. We have performed ARPES measurements at $h\nu$ = 92 eV and covered a wide {\bf k} region extending to the second $\bar{M}$ point of the surface BZ.  As illustrated in Fig. 2(b), the measured cut (blue curve) is apart from the bulk L point at the $\bar{M}$ point in the negative $k_x$ region ($\bar{M}$$_{-1}$), whereas in the positive $k_x$ region, the cut nearly crosses the L point in both the first and second BZ boundaries (indicated by $\bar{M}$$_{+1}$ and $\bar{M}$$_{+2}$, respectively). Reflecting such a difference in the $k_z$ location, the intensity at $E_{\rm F}$ [Fig. 2(c)] around the $\bar{M}$$_{-1}$ point in the FS mapping is very weak [see also the band dispersion in Fig. 2(d)] while the intensity around the $\bar{M}$$_{+1}$ and $\bar{M}$$_{+2}$ points is prominent.  This gives conclusive evidence for the presence of a small bulk FS at the L point. Note that the band dispersion along the direction parallel to the BZ boundary [cuts A-C shown in Fig. 2(c)] also supports this conclusion, because the band approaches or touches $E_{\rm F}$ only around the $\bar{M}$ point [Fig. 2(e)].
   
   The next issue to be clarified is the effect of indium substitution on the electronic states of SnTe.  As already mentioned, pristine SnTe was recently identified as a TCI \cite{TanakaNP, FuTCIPRL, FuTCINC, DziawaNM, HasanPbSnTeNC}, in which a topologically nontrivial Dirac-cone surface state appears; importantly, the {\bf k} location of the Dirac point, defined $\bar{\Lambda}$ point, is slightly away from the time-reversal-invariant $\bar{X}$ point, leading to a pair of Dirac cones to appear near $\bar{X}$ \cite{TanakaNP}.  Figures 3(a) and (b) show a direct comparison of near-$E_{\rm F}$ ARPES intensity between pristine and In-doped SnTe measured along a {\bf k} cut crossing the $\bar{\Lambda}$ point. A highly dispersive holelike band is clearly visible in pristine SnTe. This band is attributed to an admixture of the bulk and surface bands with dominant contribution from the surface state near $E_{\rm F}$, as confirmed by the $h\nu$ independence of the band dispersion for $|$${E_{\rm B}}$$|$ $<$ 0.2 eV \cite{TanakaNP}. On the other hand, the overall dispersive feature is significantly broader in In-SnTe. In fact, a direct comparison of a selected momentum distribution curve (MDC) at $E_{\rm B}$ = 0.5 eV in Fig. 3(d) and its numerical fittings suggest that the MDC width for In-SnTe, which reflects the inverse lifetime of quasiparticles ({\it i.e.} their scattering rate), is about 3 times larger than that for SnTe, in line with the transport results showing $\sim$10 times shorter lifetime in In-SnTe.
   
Nevertheless, it is still possible to extract the intrinsic band dispersion for In-SnTe despite the broad spectral feature, by taking second derivatives of the MDC as displayed in Fig. 3(c). The result of such an analysis clearly shows a linearly dispersive holelike band approaching $E_{\rm F}$, which is similar to what is found in pristine SnTe; indeed, the extracted band dispersions for SnTe and In-SnTe shown in Fig. 3(e) overlap with each other for both $h\nu$ = 21 and 92 eV within the experimental uncertainty (note that the chemical-potential difference between the two compounds is estimated to be less than 10 meV from the Sn 4$d$ core-level measurements, well within the uncertainty of the extracted band dispersion of $\pm$20 meV). This result demonstrates that the surface state is likely to be present in the vicinity of $E_{\rm F}$ in In-SnTe, even though it is significantly broadened and weakened due to the enhanced quasiparticle scattering rate. This conclusion is further supported by our observation of the ``M''-shaped band dispersion along the $\bar{\Gamma}$$\bar{X}$ cut [Fig. 3(f)], which can be regarded as a fingerprint of the Dirac-cone surface state of a TCI such as SnTe \cite{TanakaNP} and Pb$_{0.6}$Sn$_{0.4}$Te \cite{HasanPbSnTeNC} where the calculated bulk-band structure never shows any M-shaped dispersion near $E_{\rm F}$ \cite{LittlewoodPRL, HasanPbSnTeNC}.

\begin{figure}
\includegraphics[width=3.4in]{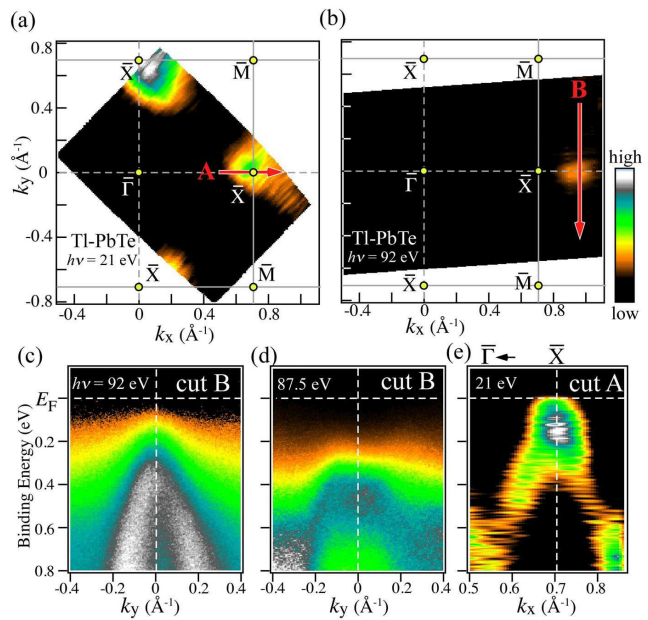}
\vspace{-0.4cm}
\caption{(Color online): (a) and (b), ARPES intensity mapping for Tl-PbTe ($x$ = 0.013) at $E_{\rm F}$ measured on the (001)-cleaved surface at $T$ = 30 K with $h\nu$ = 21 and 92 eV, respectively, plotted as a function of in-plane wave vector. (c) and (d), Near-$E_{\rm F}$ ARPES intensity as a function of $k_y$ and $E_{\rm B}$ along a {\bf k} cut across the $\bar{\Lambda}$ point [red arrow in (b)] for $h\nu$ = 92 and 87.5 eV, respectively. (e) Band dispersion derived from the second derivatives of the MDCs along the $\bar{\Gamma}$$\bar{X}$ cut for Tl-PbTe.
}
\end{figure}

   Now we turn our attention to the electronic states of Tl-PbTe which is regarded as a non-TSC counterpart of In-SnTe \cite{SasakiInPRL}. As shown in the FS mapping at $E_{\rm F}$ for the (001)-cleaved surface, there is a bright intensity spot around the $\bar{X}$ point for $h\nu$ = 21 eV [Fig. 4(a)] whose intensity center moves away from the $\bar{X}$ point at $h\nu$ = 92 eV [Fig. 4(b)]. This spot originates from a highly dispersive holelike band with a finite $k_z$ dispersion, as demonstrated in Figs. 4(c) and (d) for $h\nu$ = 92 and 87.5 eV. These observations are essentially similar to those in In-SnTe (see Fig. 1), indicating that there is a small hole pocket centered at the L point also in Tl-PbTe \cite{NakayamaTlPbTePRL}. Despite this commonality in the bulk FS topology between In-SnTe and Tl-PbTe, we found that the surface state is absent in Tl-PbTe, unlike In-SnTe. As demonstrated in Fig. 4(e), there is no signature of the M-shaped band in Tl-PbTe along the $\bar{\Gamma}$$\bar{X}$ cut. Therefore, it is most likely that the band inversion is absent in Tl-PbTe as is the case with the parent compound PbTe \cite{TanakaNP}, which means that the valence band parity is opposite between the two superconductors, In-SnTe and Tl-PbTe.
   
   Now we briefly discuss the implications of our results for the possible occurrence of topological superconductivity in In-SnTe but not in Tl-PbTe. First, the present experiments elucidate that the bulk band inversion, which makes SnTe a TCI, is kept unchanged in In-SnTe. This gives solid footing to the argument \cite{SasakiInPRL} that the pairing between Sn and Te $p$ orbitals possessing opposite parity would lead to an odd-parity SC state that can be shown to be topological; if the odd-parity pairing is indeed realized, In-SnTe is a 3D TSC. On the other hand, the absence of band inversion in Tl-PbTe suggests that Pb and Te $p$ orbitals relevant at the Fermi level have the same parity, which precludes the above mechanism for odd-parity pairing.  Second, the existence of the topological surface state at $E_{\rm F}$ can be responsible for unusual SC properties in In-SnTe, because the opening of the bulk SC gap may lead to an opening of a separate, proximity-induced SC gap in the topological surface state \cite{HasanCuNP, HasanReview, SCZhangReview}; in this case, even if the bulk SC is a conventional BCS type, the surface may host a 2D TSC. The proximity-induced superconductivity in the topological surface state of a TCI has not yet been addressed, and its consequence, including whether Majorana fermions can be conceived there, is an interesting future topic. Our observation of the topological surface state in In-SnTe in the normal state assures that this material provides a fertile ground for the study of TSC irrespective of whether the putative odd-parity pairing is realized in the bulk. 
   
   In summary, we have performed ARPES experiments on In-SnTe and Tl-PbTe to elucidate the energy-band structures underlying their SC states. We could barely resolve the topological surface state characteristic of a TCI in In-SnTe despite the significant broadening of the spectra due to enhanced quasiparticle scatterings caused by In-doping. This gives evidence for an inverted band structure as in pristine SnTe and points to interesting physics associated with either the odd-parity pairing in the bulk or the proximity-induced superconductivity in the surface state of a doped TCI below $T_{\rm c}$, which has not yet been addressed. In contrast, in Tl-PbTe we found no surface state, probably due to the absence of band inversion in this compound. These results are in good correspondence with the recent point-contact spectroscopy experiment \cite{SasakiInPRL} which found pronounced ZBCP in In-SnTe but only conventional Andreev reflection in Tl-PbTe.

\begin{acknowledgments}
We thank Liang Fu for stimulating discussions. We also thank M. Nomura, T. Shoman, K. Honma, H. Kumigashira, K. Ono, M. Matsunami, S. Kimura, for their assistance in ARPES measurements, and T. Ueyama and K. Eto for their assistance in crystal growth. This work was supported by JSPS (NEXT Program and KAKENHI 23224010), JST-CREST, MEXT of Japan (Innovative Area gTopological Quantum Phenomenah), AFOSR (AOARD 124038), KEK-PF (Proposal number: 2012S2-001), and UVSOR (Proposal number: 24-536).
\end{acknowledgments}

\end{document}